\documentclass[journal=jacsat,manuscript=article]{achemso}
\pdfoutput=1
\usepackage{chemformula}
\usepackage[T1]{fontenc}
\usepackage{achemso,array,booktabs,lmodern}
\usepackage[osf]{mathpazo}
\usepackage[scaled=0.95]{helvet}
\usepackage[final]{listings,microtype}
\usepackage[T1]{fontenc}
\usepackage{caption}
\usepackage{subcaption}
\usepackage{float}
\usepackage{geometry}
\usepackage{natbib}
\usepackage{setspace}
\usepackage{xkeyval}
\usepackage{float}
\usepackage[symbol]{footmisc}
\author{Vahid Nikkhah}
\affiliation{University of Pennsylvania, Department of Electrical and Systems Engineering, Philadelphia, PA 19104 U.S.A.}
\author{Mario Junior Mencagli} 
\affiliation{Department of Electrical and Computer Engineering, University of North Carolina, Charlotte, NC 28223, USA}
\author{Nader Engheta}
\email{engheta@ee.upenn.edu}
\affiliation{University of Pennsylvania, Department of Electrical and Systems Engineering, Philadelphia, PA 19104 U.S.A.}

\title
    {Reconfigurable nonlinear optical element using tunable couplers and inverse-designed structure}

\begin{document}
\begin{abstract}
In recent years, wave-based analog computing has been at the center of attention for providing ultra-fast and power-efficient signal processing enabled by wave propagation through artificially engineered structures. Building on these structures, various proposals have been put forward for performing computations with waves. Most of these proposals have been aimed at linear operations, such as vector-matrix multiplications.
 The weak and hardly controllable nonlinear response of electromagnetic materials imposes challenges in the design of wave-based structures for performing nonlinear operations. In the present work, first, by using the method of inverse design we propose a three-port device, which consists of a combination of linear and Kerr nonlinear materials, exhibiting the desired power-dependent transmission properties. Then, combining a proper arrangement of such devices with a collection of Mach-Zehnder interferometers (MZIs), we propose a reconfigurable nonlinear optical architecture capable of implementing a variety of nonlinear functions of the input signal. The proposed device may pave the way for wave-based reconfigurable nonlinear signal processing that can be combined with linear networks for full-fledged wave-based analog computing.
\end{abstract}

\section{Introduction}
Optical nonlinearity plays a crucial role in several exciting areas of research\cite{guo2016chip,jiang2016analog,lapine2014colloquium,krasnok2018nonlinear}. 
Among these areas, emerging optical neural networks (ONNs)\cite{shen2017deep,hamerly2019large} which are a physical implementation of a standard artificial neural network (ANN) with optical components, have received growing interest.
ONNs have several advantages over their electronic counterparts.
Among these advantages, it is worth emphasizing their higher computational speed with lower power consumption.
These features make them appealing for several applications that require handling large data sets such as real-time image processing\cite{krizhevsky2012imagenet}, language translation\cite{young2018recent}, decision-making problems\cite{najafabadi2015deep}, and more\cite{silver2018general,wang2016deep}.
One of the most important units in the ANN structure is the activation function\cite{karlik2011performance}.
This function determines the neural network output, its accuracy, and the computational efficiency of a training model.
Activation functions can be described by simple mathematical nonlinear functions.
In recent years, different nonlinear functions, acting as activation functions, have been investigated in the ANN community.
It turned out that the nonlinear activation function's choice is closely connected with the ANN application\cite{karlik2011performance}.
Diverse ANN applications require the use of different nonlinear activation functions.
For example, the sigmoid function, which is the most common nonlinear activation function, is particularly suitable for applications that produce output values in the range of $[0,1]$.
Despite the mathematical simplicity of nonlinear activation functions, it is still unclear how to perform arbitrary nonlinear functions on waves physically or even in principle whether such functions are generally possible.
Current nonlinear optical components, such as bistable and saturable absorber devices\cite{shen2017deep,hamerly2019large,harris2018linear}, can implement only a subset of all the possible nonlinear activation functions.
Moreover, most of them suffer from a lack of reconfigurability, which means that once they have been realized, the form of their nonlinear response cannot be changed.
As a result, they can only be used to implement a single activation function limiting the ONN application range.
In the present work, we introduce an idea for a \emph{reconfigurable} architecture that can implement arbitrary nonlinear functions (within certain constraints) between the input signal, and the output signal (See Figure 1).
The proposed architecture is composed of two networks: (1) the nonlinear signal divider and (2) the linear optical network composed of a mesh of Mach-Zehnder Interferometers (MZIs).
The nonlinear optical network consists of a set of identical three-port devices functioning as specialized optical power limiters.
These limiters employ a Kerr nonlinear material and a linear dielectric for limiting the optical signal/intensity of the light that is coupled to one of the output ports and the rest of the energy is coupled to another output port.
The specific composition and spatial distributions of the Kerr material and the linear dielectric are obtained through the method of the inverse design\cite{bendsoe2003topology,lu2011inverse,piggott2015inverse,piggott2017fabrication,molesky2018inverse,hughes2018adjoint}.
This method has opened up enormous opportunities both in linear and nonlinear optics for designing nonintuitive optical devices with complex functionalities.
The optically-linear MZI portion of the proposed device consists of a properly arranged collection of MZIs, as shown in Figure 1.
Photonic MZI meshes, which commonly consist of waveguide-based MZI array laid out on a flat silicon substrate, have attracted a great deal of attention in recent years\cite{harris2018linear,miller2015sorting,MIllerJLT,MillerOptExp,taballione20198,bogaerts2020programmable,harris2017quantum}.
This interest mostly focuses on MZI networks that can be programmed to implement any linear transformations with electromagnetic waves\cite{reck1994experimental,miller2013self,clements2016optimal}.
Since such networks exploit the well-known computational capabilities of electromagnetic waves and are suitable for on-chip integration, it is playing an important role in several applications, including optical machine learning\cite{shen2017deep}, quantum-information processing\cite{harris2017quantum}, forward scattering problems\cite{nikkhah2022inverse}, and optical analog equation solving platform\cite{mencagli2018solving,tzarouchis2022mathematical}. 
Our proposed device here, thanks to its versatility and reconfigurability, may be useful for various applications in nonlinear photonics, such as ONNs mentioned above, in which reconfigurable nonlinear optical elements are needed.

Throughout the paper, we assume $e^{j \omega t}$ as the time dependence of EM fields.
\begin{figure}[!ht]
    \centering
    \includegraphics[width = \textwidth]{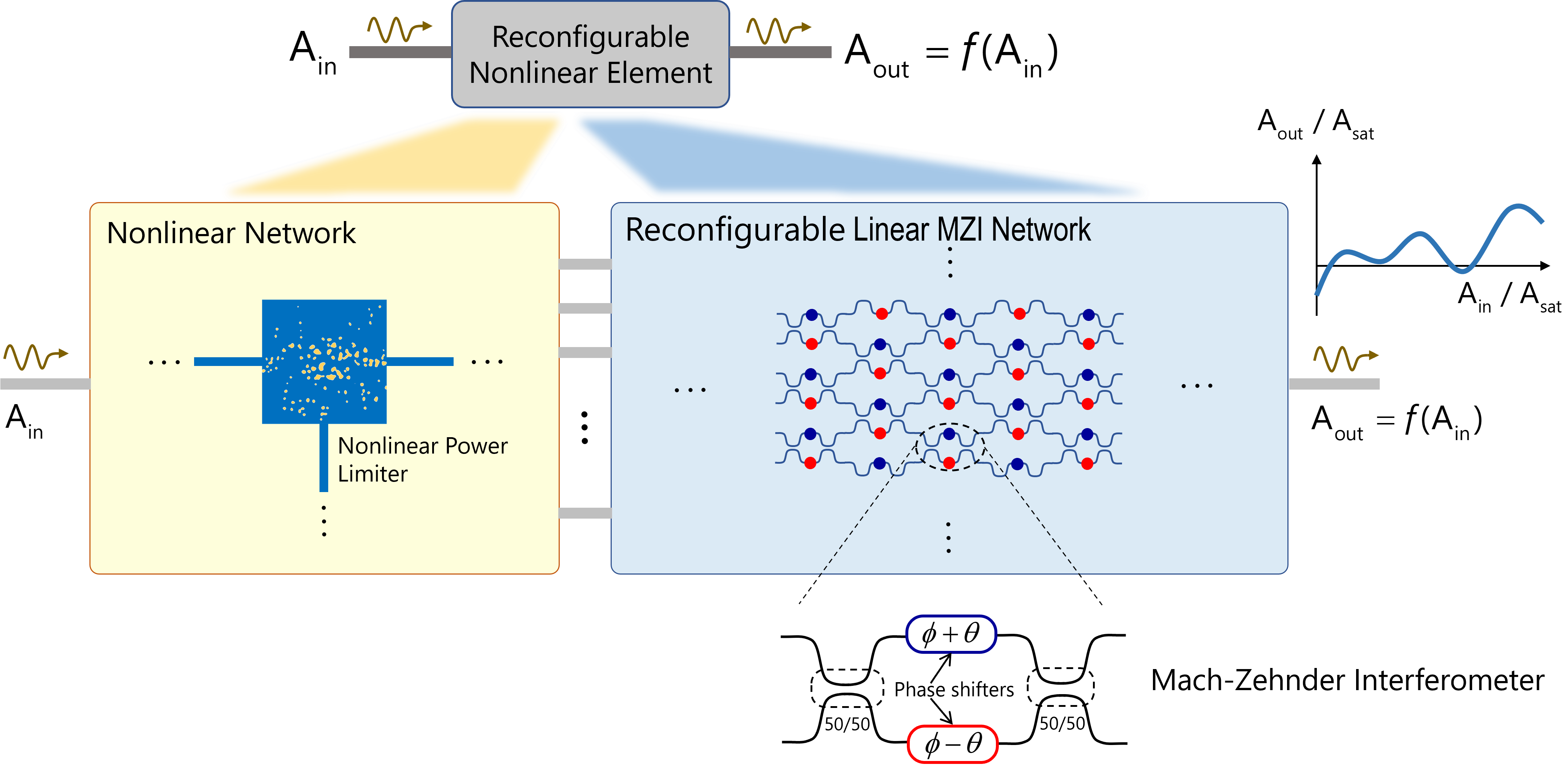}
    \caption{General reconfigurable optical architecture for generating tunable nonlinear functions. The architecture is composed of a nonlinear and a linear network. The nonlinear network (yellowish-filled rectangular box) consists of a set of identical nonlinear three-port devices, each of which acts as a nonlinear power limiter. The linear network (bluish-filled rectangular box) consists of a collection of linear MZIs. The inset sketches a generic MZI mesh configuration.}
    \label{proposed_architecture}
\end{figure}
\section{Inverse-designed nonlinear power limiter for the nonlinear network}
\begin{figure}[!ht]
    \centering
    \includegraphics[width =\textwidth]{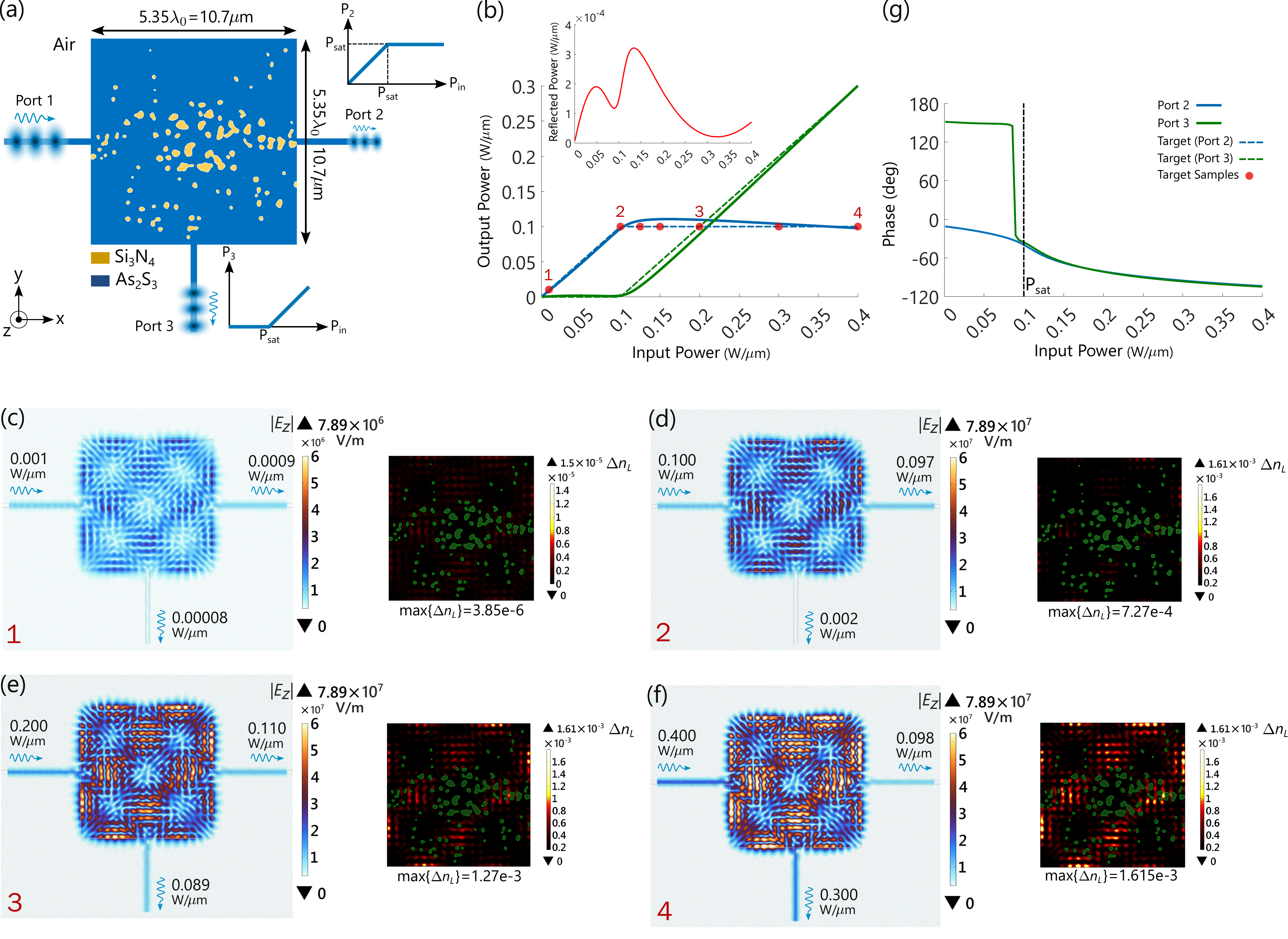}
    \caption{\textbf{Inverse-designed nonlinear photonic structure as power limiter.}(a) A transverse-electric (TE)-polarized optical signal is the input to the waveguide denoted as Port 1. The power of the output signal at Port 2 follows the power of the input signal but is limited when the input power increases beyond $P_\text{sat}$.  The rest of the energy is then directed toward Port 3.
    The optimized distribution of the Kerr nonlinear material ($\text{As}_2\text{S}_3$) and the linear dielectric ($\text{Si}_3\text{N}_4$) are, respectively, shown by blue and yellow regions.
    (b) The blue and green solid curves, respectively, show the transmitted power of the signals at Port 2 and Port 3 of the proposed element in panel (a) vs. the input power. The dashed curves are the corresponding desired transmission plots. The red solid curve illustrates the reflected power back to Port 1. The eight red circles are the target samples of the input power for which the cost function is defined for optimization. (c)-(f) The magnitude of the electric field distribution (left side of each panel) and the nonlinear refractive index shift inside the Kerr medium (right side of each panel) simulated for different input power values, $P_\text{in} = 0.01P_\text{sat},P_\text{sat},2P_\text{sat},4P_\text{sat}$. (g) The phases of the output signals at Port 2 (blue solid curve) and Port 3 (green solid curve) relative to the input phase.}
    \label{Des_Sim}
\end{figure}

In this section, the required characteristics and design of the constitutive element of the nonlinear network, i.e., the nonlinear power limiter are discussed.
As schematically illustrated in Fig.~\ref{Des_Sim}(a), the nonlinear element is a three-port device that is required to exhibit the following transmission features.
When a monochromatic signal, which is injected into the input port (Port 1) carries a power below a certain value, denoted as $P_\text{sat}$, this signal should appear at the designated output port on the right (Port 2).
If the power of the input signal exceeds $P_\text{sat}$, the power of the signal at Port 2 should stay at its "saturated" value,$P_\text{sat}$, and the spillover of the input signal should be directed to the other output port on the bottom (Port 3).
As the power of the input signal increases beyond $P_\text{sat}$, the power reaching Port 3 keeps increasing, while the one going to Port 2 remains constant and equal to $P_\text{sat}$. 
Also, as will become clear shortly, it is required that the phases of the output signals at Ports 2 and 3 be the same (and, with an additional part discussed in the MZI section, to be the same as the phase of the input signal).
To achieve these transmission characteristics, we consider a design region containing a distribution of Kerr nonlinear and linear dielectric materials connected to the input and output ports through slab waveguides as depicted in Fig.~\ref{Des_Sim}(a).
The cladding surrounding the structure is assumed to be air.
Without loss of generality, we also assume the structure to be two-dimensional (2D) meaning it is infinitely extended in the out-of-plane direction ($z$ axis). 
The materials of the structure are invariant along the out-of-plane coordinate ($z$) and, as a result, the distribution of the electromagnetic (EM) fields is computed only over the in-plane geometry ($x$-$y$ plane).
Our proposed approach here is general and the choice of materials depends on the frequency of operation. 
Here, as a nonlinear material, we used a realistic high-index Kerr material such as $\text{As}_2 \text{S}_3$ (Arsenic sulfide), which is modeled with the following intensity-dependent permittivity\cite{boyd2020nonlinear}:
\begin{equation}
	\varepsilon_r^\text{NL} = \varepsilon_r^\text{L} + 3 \chi^{(3)} |\textbf{E}|^2
\end{equation}
where $\varepsilon_r^\text{L}$ and $\chi^{(3)}$ denote the linear relative permittivity and the Kerr coefficient, respectively. At an operating free-space wavelength of $\lambda_0 = 2 \ \mu m$, $\text{As}_2 \text{S}_3$ presents a linear permittivity of $\varepsilon_r^\text{L} = 5.76$, and a relatively high Kerr coefficient of $\chi^{(3)} = 4.1 \times 10^{-19} \ \text{V}^2 / \text{m}^2$.
As shown in Fig.~\ref{Des_Sim}(a), the slab waveguides' cores are assumed to be $\text{As}_2 \text{S}_3$ with a width of $\lambda_0 / 6$. 
This width only allows the propagation of the fundamental TE (transverse electric) mode [$\textbf{E}=\left(0,0,E_z\right)$]. Within the design region, the distribution of $\text{As}_2 \text{S}_3$ and a linear dielectric material such as $\text{Si}_3 \text{N}_4$ (Silicon nitride with $\varepsilon_r^{\text{Si}_3 \text{N}_4} = 4$ at $\lambda_0 = 2\mu m$) is optimized to achieve the desired transmission features.
Note that this choice of materials is just an example and other materials can also be used.
For example, at $\lambda_0  = 1.55 \ \mu m$ CMOS-compatible a-SiC (amorphous silicon carbide)\cite{xing2019cmos} and $\text{Si} \text{O}_2$ (silica) can also be considered for the design ,respectively, as the Kerr and linear dielectric materials.
The distribution of the two materials ($\text{As}_2 \text{S}_3$ and $\text{Si}_3 \text{N}_4$) in the design region is obtained by the density-based topology optimization technique known as the inverse design method \cite{lu2011inverse,piggott2015inverse,piggott2017fabrication,molesky2018inverse}.
The goal of the optimization process is to determine the proper distribution of the two materials to provide the device with the transmission characteristics discussed above with a priori chosen saturation power e.g.,$P_\text{sat} = 0.1 \ \text{W}/ \mu \text{m}$. 
(The choice of $P_\text{sat}$ value here is an example, and in general, this value depends on the choice of the Kerr nonlinear material. Moreover, since we are dealing with a two-dimensional structure, the power is given as $ \text{W}/ \mu \text{m}$ where $\mu \text{m}$ is along the z-axis. We choose $P_\text{sat} = 0.1 \ \text{W}/ \mu \text{m}$ which provides large enough refractive index shift inside $\text{As}_2 \text{S}_3$ for our proposes and it is small enough that maintains a safe margin from damaging the materials within the design region.)
Namely, the power injected at Port 1 is coupled to Ports 2 and 3 according to the blue and green dashed curves, respectively, as shown in Fig.~\ref{Des_Sim}(b).
The optimization procedure, which is discussed in details in supplementary information, was carried out to match the transmission from Port 1 to Ports 2 and 3 at a discrete number of input power values (see the red circles in Fig.~\ref{Des_Sim}(b)), resulting in the optimized distribution of $\text{As}_2 \text{S}_3$ and $\text{Si}_3 \text{N}_4$ shown in Fig.~\ref{Des_Sim}(a).
The distribution of the magnitude of electric field ($|E_z|$) and the nonlinear refractive index shift ($\Delta n_\text{NL} = \sqrt{\varepsilon_r^\text{L}+3\chi^{(3)}|E_z|^2}-\sqrt{\varepsilon_r^\text{L}}$) that are simulated for the optimized device at four different input power values are shown in Fig.~\ref{Des_Sim}(c)-(f).
For instance, Fig.~\ref{Des_Sim}(c) shows $|E_z|$ and $\Delta n_\text{NL}$ for $P_\text{in} = 0.01 P_\text{sat} = 0.001 \ \text{W}/ \mu \text{m}$ which is well below the saturation power. 
With this level of input power, the nonlinear response of Kerr material induced by the local field intensity is very weak.
The maximum observed nonlinear index shift inside the Kerr medium is $3.85 \times 10^{-6}$ (see the right panel), which is basically negligible, and as a result, the device operates in the linear regime. 
As can be seen on the left panel, for $P_\text{in} = 0.01 P_\text{sat}$, most of the input power is coupled to Port 2, $P_2 = 0.0009 \ \text{W}/ \mu \text{m}$, as desired. Only a small amount of leakage is observed at Port 3, $P_3 = 0.00008  \ \text{W}/ \mu \text{m}$.
By increasing the input power up to the edge of the saturation i.e., $P_\text{in} = P_\text{sat}$, the nonlinear response of the Kerr material kicks in, as can be observed on the right panel of Fig.~\ref{Des_Sim}(d). 
The maximum shift of the refractive index is $7.27 \times 10^{-4}$, which is two orders of magnitude larger than the one observed with $P_\text{in} = 0.01 P_\text{sat}$ (see the right panels of Fig.~\ref{Des_Sim}(c) and (d)).
Nevertheless, most of the input power is still coupled to Port 2 ($P_2 = 0.097 \ \text{W}/ \mu \text{m}$) as desired and there is a small leakage of $P_3 = 0.002 \ \text{W}/ \mu \text{m}$ at Port 3. 
Also, a negligible portion of the input power equal to $P_\text{in} - (P_2 + P_3) = 0.001 \ \text{W}/ \mu \text{m}$ is either scattered away from the device to the air cladding or reflected back towards the input port, as shown in the inset of Fig.~\ref{Des_Sim}(b). 
By bringing the input power above the saturation level, i.e., $P_\text{in} = 2P_\text{sat}$, the nonlinear response becomes more pronounced as indicated by the right panel of Fig.~\ref{Des_Sim}(e) showing a maximum $\Delta n_\text{NL} = 1.27 \times 10^{-3}$. 
As can be observed in the right panel of Fig.~\ref{Des_Sim}(e), the collective power-dependent response is strong enough to limit  the power at Port 2 ($P_2 = 0.110 \ \text{W}/ \mu \text{m}$) around the level of the saturation power ($P_\text{sat} = 0.100 \ \text{W}/ \mu \text{m}$).
The remaining power, $P_3 = 0.089 \ \text{W}/ \mu \text{m}$, appears at Port 3, as expected.
As the input power increases further beyond $P_\text{sat}$, i.e., $P_\text{in} = 4P_\text{sat}$, the increase of the nonlinear effect enables maintaining the power at Port 2 almost constant around $P_\text{sat}$ and directing the extra power at Port 3 (See Fig.~\ref{Des_Sim}(f)).

So far, we have discussed the performance of the three-port inverse-designed nonlinear power limiter for a discrete number of input power values.
In Fig.~\ref{Des_Sim}(b), the solid blue and green lines, respectively, show the transmitted power from Port 1 to Ports 2 and 3 which are calculated by continuously sweeping the input power from $0$ to $4P_\text{sat}$.
As observed from the transmission plots, The agreement with the target responses (dashed curves) is quite good.
For $P_\text{in} \leq P_\text{sat}$, the power reaching Port 2 increases linearly.
For the input powers beyond $P_\text{sat}$, the output power at Port 2 saturates around $P_\text{sat}$ and the power going to Port 3 starts increasing almost linearly with $P_\text{in}$.
Fig.~\ref{Des_Sim}(g) shows the relative phase variation between the output signals (Ports 2 and 3)  and the input signal (Port 1) versus the input power.
For $P_\text{in} < P_\text{sat}$, the phase at Port 3 is immaterial, as the amount of power reaching this port is very small.
For $P_\text{in} > P_\text{sat}$, the phase of the signals at Ports 2 and 3 assumes a similar behavior decreasing quasi-linearly as the input power increases.
This phase variation is expected with Kerr nonlinear materials with a positive $\chi^{(3)}$.
An increase in the input power implies an increase in the refractive index resulting in a decrease in the output signal phase (note that we are using $e^{j \omega t}$ as the time dependence).
Considering that a properly arranged set of these inverse-designed nonlinear three-port elements feed the MZI mesh that requires proper phase relations for its input, the phase behavior of the three-port output signals needs to be properly adjusted for the MZI mesh.
This issue will be addressed in the following section.
The MZI mesh works with complex-valued signals ($A$) containing both magnitude and phase information. The phases of the output signals leaving the inverse-designed nonlinear element are available in Fig.~\ref{Des_Sim}(g).
Their magnitudes ($|A|$) can be retrieved from the output power values (See Fig.~\ref{Des_Sim}(b)) through the expression $P = \frac{1}{2Z_t} \int |A \psi(l)|^2 dl$.
$Z_t$ denotes the transverse impedance of the fundamental TE mode supported by the input/output waveguides (For the lossless waveguide, which we assumed here, this transverse impedance is real-valued.) 
$\psi(.)$ is the eigenmode profile that is normalized such that $\int |\psi(l)|^2 dl = 1$, with $l$ denoting the coordinate along a slab waveguide's cross-section.
Note that, in doing so, $\psi(.)$ is dimensionless whereas $A$ has the dimension of the electric field [$V/m$]. 
The magnitude of the signals leaving the nonlinear element can be calculated from the power as $|A|= \sqrt{2Z_tP}$.
\section{Adding Reconfigurable MZI network}
\begin{figure}[!ht]
    \centering
    \includegraphics[width = \textwidth]{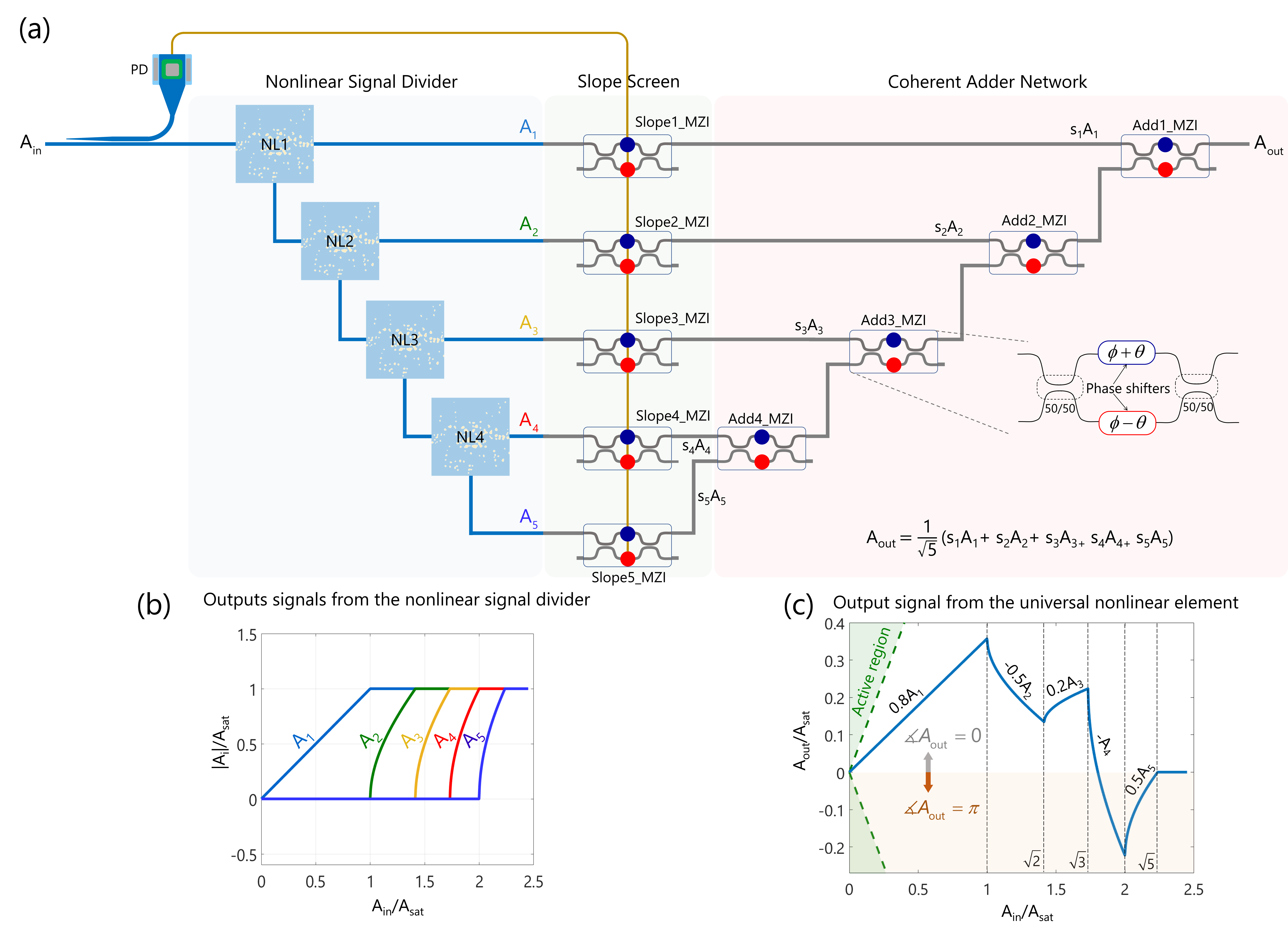}
    \caption{Schematic illustration of our proposed nonlinear architecture composed of a nonlinear signal divider (NSD) and a linear network of MZIs. The 1-to-N nonlinear signal divider is a cascaded architecture of N-1 individual nonlinear power limiters that distribute the input signal with amplitude $A_\text{in}$ to N output ports in a particular way, as the input power increases. Two MZI sets identified as "slope screen" and "coherent adder network" are the sub-parts of the linear MZI network. The slope screen scales the outputs of NSD each with a coefficient $-1 \leq s \leq 1$. The adder network adds the scaled signals with an equal weight of $1\sqrt{N}$ to generate the output signal with amplitude $A_\text{out}$. A tap waveguide connected to a photodetector produces a power-dependent voltage signal that modifies the common phases of the MZIs in the slope screen for compensating the phases of the nonlinear limiters. (b) The amplitudes of the signals at the 5 outputs of NSD. These signals form a set of basis functions for generating a tunable function of the input signal i.e., $A_\text{out} = f(A_\text{in})$ over a specific range of the input power. (c) An example of the output signal as a function of the input signal for the slopes $s=[0.8,-0.5,0.2,-1,0.5]$ of the slope-MZIs that are set from top to bottom. A negative output means a $\pi$ phase shift relative to the input.}
    \label{universal_nonlinear_element_details_results}
\end{figure}

In the previous section, the design of the nonlinear power limiter with the specific relationship between the input and the two output signals was presented. 
As shown in Fig.~\ref{universal_nonlinear_element_details_results}(a), we cascade $N-1$ such nonlinear elements with identical functionality into a special 1-to-N nonlinear signal divider (NSD). 
Then we feed the outputs of the NSD to a properly arranged collection of MZIs.
The MZI network processes and combines the complex-valued signals coming from the NSD to provide an output signal that is a tunable nonlinear function of the input signal i.e., $A_\text{out} = f(A_\text{in})$. 
Without loss of generality, we consider $A_\text{in}$, which denotes the input signal to the proposed architecture, to be a real positive number.
On the other hand, the output signal ($A_\text{out}$) can be a positive or negative real number. 
As will be clear shortly, the proposed architecture enables generating an output signal ($A_\text{out}$) with 0 or $\pi$ phase despite the positive input signal, which is a feature to implement a large set of nonlinear functions.
Now, let us discuss the functionality of the proposed reconfigurable nonlinear architecture beginning with the nonlinear signal divider.
As can be seen in Fig.~\ref{universal_nonlinear_element_details_results}(a), Port 3 of each nonlinear element is connected to the input port of the next one, and Ports 2 are considered the outputs of the NSD.
As the magnitude of the input signal ($A_\text{in}$) increases from zero, Port 2 of the first nonlinear element gets activated, and the magnitude of its signal increases up to the saturation level $|A_\text{sat}|$, which is the magnitude corresponding to $P_\text{sat}$.
As $A_\text{in}$ goes beyond $|A_\text{sat}|$, the spillover goes to Port 3, which is connected to the input port of the second nonlinear element. 
Therefore, Port 2 of the second nonlinear element gets activated next.
This process continues until the magnitudes of all the output signals are saturated at $|A_\text{sat}|$.
For instance, let us assume the magnitude of the input signal is between $\sqrt{2} |A_\text{sat}| \leq A_\text{in} \leq \sqrt{3} |A_\text{sat}|$ (corresponding to $2P_\text{sat} < P_\text{in} < 3P_\text{sat}$). 
Within this interval, the magnitudes of the signals at the first two output ports of the NSD are fixed at $|A_\text{sat}|$, whereas the difference between the input power and the power of the saturated signals appears at the third output port i.e., $A_3 = \sqrt{|A_\text{in}|^2 - 2|A_\text{sat}|^2}$.
In general, for the input amplitude in the range $\sqrt{m-1}|A_\text{sat}| \leq |A_\text{in}| \leq \sqrt{m} |A_\text{sat}| \ (m \leq N)$, the magnitudes of the outputs $A_k$, with $k=1,2,...,m-1$, are fixed at $|A_\text{sat}|$, the magnitude of the $m^\text{th}$ output signal is $A_m = \sqrt{|A_\text{in}|^2-(m-1)|A_\text{sat}|^2}$, and the rest are zero, $A_k=0, \ \ k=m+1,m+2,...,N$.
Fig.~\ref{universal_nonlinear_element_details_results}(b) shows the magnitude of the first five output signals as a function of $A_\text{in}$. 
The signals leaving the NSD can be seen as a set of nonlinear basis functions that are properly processed upon entering the MZI mesh to generate a continuous nonlinear output signal ($A_\text{out}$) as a reconfigurable function of the input signal ($A_\text{in}$).

\begin{figure}[!ht]
    \centering
    \includegraphics[width = \textwidth]{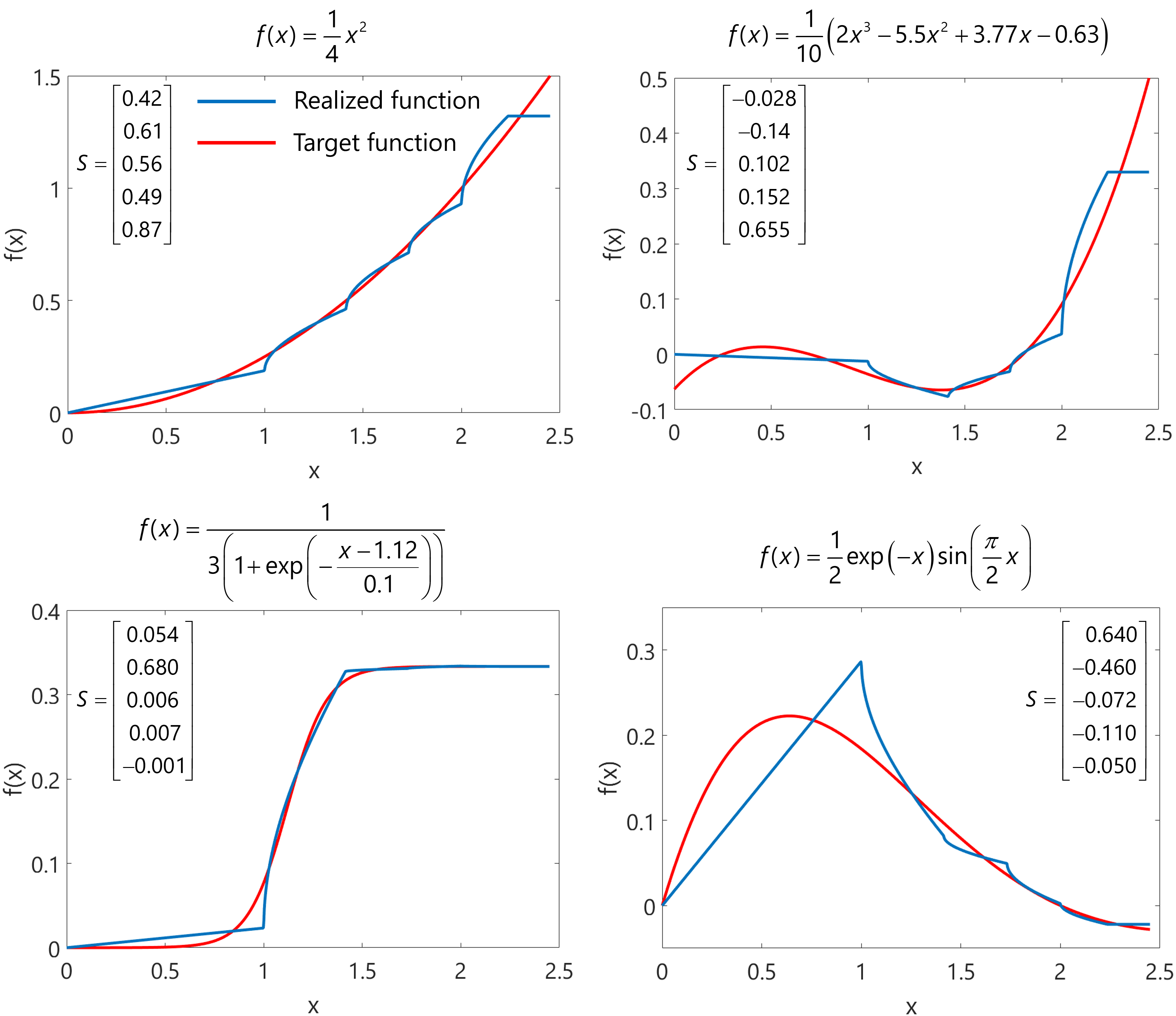}
    \caption{Examples of different target nonlinear functions of the input signal by our proposed reconfigurable nonlinear element shown in Figure2(a). Red curves are the target functions of the input signal the expression of which are shown at top of the plots. The blue curves are the resulting functions from our proposed element, by optimizing the scales of the slope screen using minimization of the MSE error between target and realized functions. The inset in each panel shows the optimized scales of the corresponding realized function.}
    \label{nonlinear_funcs}
\end{figure}

Each MZI consists of two $50\%$ beam splitters and two phase shifters parametrized by $\theta$, the differential phase, and $\phi$, the common phase\cite{shen2017deep,harris2018linear,miller2015sorting,miller2013self}, as shown in the inset at the bottom of Fig.~\ref{universal_nonlinear_element_details_results}(a). 
In silicon photonics, beam splitters are usually realized by directional couplers that transform complex-valued input signals $\textbf{a}_1$ and $\textbf{a}_2$ into complex-valued output signals $\textbf{b}_1$ and $\textbf{b}_2$ according to $\textbf{b}_{1,2} = \frac{1}{\sqrt{2}} \left ( \pm j \textbf{a}_{1,2} + \textbf{a}_{2,1} \right )$\cite{miller2013self}.; Assuming lossless beam splitters and phase shifters, the MZI implements a $2\times2$ unitary transformation between the input and output signals that can be mathematically described by the following transmission matrix\cite{miller2013self}:
\begin{equation}\label{Tmzi}
    \textbf{T}_\text{MZI} =
    \begin{pmatrix}
    	-j \sin({\theta}) e^{j\phi} & -j \cos({\theta}) e^{j\phi} \\
        -j \cos({\theta}) e^{j\phi} & j \sin({\theta}) e^{j\phi}
    \end{pmatrix}
\end{equation}
As depicted in Fig.~\ref{universal_nonlinear_element_details_results}(a), the linear network of MZI mesh can be broken up in two sub-networks: (1) the MZI collection that sets the "slopes" (denoted as the “slope screen” in Fig.~\ref{universal_nonlinear_element_details_results}(a)), which are the coefficients that would be multiplied by the $A_m = \sqrt{|A_\text{in}|^2-(m-1)|A_\text{sat}|^2}$, and (2) the MZI collection that coherently combines the signals (depicted as the “adder” section).
Namely, it performs the addition of $N$ input signals into one output signal.
In setting up the phases of each MZI to achieve such adding functionality, the phases of the $N$ input signals are assumed to be in phase.
The phase shifters of those MZIs are set progressively from the top to the bottom when one is to run the adder backward: sending a signal from the output port and imposing the power carried by that signal to be equally divided and appear at the input ports with the same phase.\cite{miller2013self,MillerOptica}.
As discussed in the previous section, the input and outputs of each nonlinear power limiter experience a phase difference that decreases quasi-linearly with the input power as shown in Fig.~\ref{Des_Sim}(g). 
Cascading the nonlinear power limiters as shown in Fig.~\ref{universal_nonlinear_element_details_results}(a) implies that the output phases of the NSD also decreases quasi-linearly with the input power but with a slope that increases progressively with the number of activated nonlinear elements. 
For example, let us assume the power of the input signal is within the range such that the first two nonlinear elements are activated. 
The phase of the first NSD output ($A_1$) will decrease quasi-linearly with the input power as shown in Fig.~\ref{Des_Sim}(g). 
The phase of the second output ($A_2$) will decrease as $A_1$ but with a double slope because the second nonlinear limiter is fed with a signal whose phase profile is identical to the one of $A_1$. 
A similar argument can be repeated for the other nonlinear elements. 
To compensate for this phase difference and equalize the phases of the NSD outputs with the phase of the input signal ($A_1$) as required by the adder, a voltage-driven phase shifter\cite{shen2017deep,harris2018linear,harris2017quantum} can be used to change the common phases of the MZIs in the slope screen based on the power of the input signal. 
Specifically, on the input side of the proposed architecture (See Fig.~\ref{universal_nonlinear_element_details_results}(a)), a tap waveguide connected to a photodetector producing a voltage proportional to the input power can be used to modify the common phases of the MZIs to compensate for the phase variation of the NSD outputs.
As an aside, it is worth mentioning that for the sake of simplicity and just to describe the concept, we assume that the mode propagation in all waveguides in each MZI experiences the same phase shift when they get to the adder network.
If that is not the case the phase differences due to the different lengths of these waveguides can be easily compensated for by the common phase shifter in each MZI.
Following this approach, the common phase of $m^\text{th}$ slope-MZI is set to $\phi_m (P_\text{in})= \pi / 2 + m \Delta \psi ( P_\text{in})$ where $\Delta \psi (P_\text{in})$ is the required phase adjustment produced by the photodetector signal to make the first NSD output ($A_1$) coherent with the input signal ($A_\text{in}$).
After setting the common phases according to $\phi_m (P_\text{in})$, the differential phase ($\theta_m$) determines the coupling between the $m^\text{th}$ NSD output and the $m^\text{th}$ input to the adder network according to $s_m=\sin(\theta_m)$ with $s_m$ ($\theta_m$) ranging from $-1$ ($-\pi/2$) to $1$ ($\pi/2$) (see the first entry of the matrix in Eq.~(\ref{Tmzi})).
The adder combines the signals that are leaving the slope screen into the output signal, which will be of the form $A_\text{out} = \frac{1}{\sqrt{N}} \sum_{m=1}^{N} s_\text{m} A_\text{m}$. 
The details regarding the design of the adder network are provided in the supplementary information.
As can be observed from the previous expression, the differential phase $\theta_m$ sets the slope of the portion of $A_\text{out}$ for the input amplitude between $\sqrt{m-1}A_\text{sat}$ and $\sqrt{m}A_\text{sat}$. 
By tuning these differential phases of the MZIs in the slope screen, the architecture depicted in Fig.~\ref{universal_nonlinear_element_details_results}(a) can approximately provide a variety of different nonlinear transformations, $A_\text{out} = f(A_\text{in})$, through a continuous piecewise function (for any multiplicative coefficient  between -1 and +1 before the $A_{m}$, and then multiplied by $1/\sqrt{N}$), defined on intervals of input magnitude $[\sqrt{m-1}A_\text{sat},\sqrt{m}A_\text{sat}]$.
Fig.~\ref{universal_nonlinear_element_details_results}(c) shows an example of a nonlinear function realized with the proposed architecture assuming $N=5$.
For this example, the differential phases of the slope screen are arbitrarily chosen to be $ \textbf{s} = [0.8,-0.5,0.2,-1,0.5]$.
During the first interval of the input magnitude, corresponding to ($[0,A_\text{sat}]$), all the signals leaving the NSD are zero except for $A_1$, and, as a result, the output signal is given by $A_\text{out} = \frac{1}{\sqrt{5}} \textbf{s}(1) A_\text{1}$.
Since the first MZI of the slope screen sets a positive slope ($\textbf{s}(1)=0.8$), the output signal increases with $A_\text{in}$ as can be observed in Fig.~\ref{universal_nonlinear_element_details_results}(c).
When $A_\text{in}$ goes beyond $A_\text{sat}$, entering in the second interval of the input magnitude ($[A_\text{sat},\sqrt{2}A_\text{sat}]$), the second nonlinear limiter in the NSD is activated in addition to the first one.
The output signal is now given by $A_\text{out} = \frac{1}{\sqrt{5}} (\textbf{s}(1) A_\text{1}+\textbf{s}(2) A_\text{2})$.
As discussed above, the second nonlinear limiter in the NSD is activated when the output of the first one ($A_1$) is saturated, that is, $|A_1|$ is constant with respect to the input power.
Thus, the slope of the output during the interval $[A_\text{sat},\sqrt{2}A_\text{sat}]$ is dictated by the differential phase ($\theta_2$) of the second MZI of the slope screen.
In the example under consideration, $\theta_2$ was selected to be $-\pi/6$ and the output signal decreases with the slope $\textbf{s}(2)=-0.5$ when the input magnitude goes from $A_\text{sat}$ to $\sqrt{2}A_\text{sat}$ as illustrated in Fig.~\ref{universal_nonlinear_element_details_results}(c).
A similar argument can be repeated when the input power increases further activating all the five nonlinear limiters in the considered example.
As can be observed in Fig.~\ref{universal_nonlinear_element_details_results}(c), interestingly, the proposed architecture can generate positive and negative nonlinear output signals enabling the implementation of a large set of nonlinear functions.
As the proposed architecture does not include any active components, the output signals are always confined in the region bounded by the bisector of the first and fourth quadrants of $A_\text{out} - A_\text{in}$ Cartesian diagram (See Fig.~\ref{universal_nonlinear_element_details_results}(c)).
We consider four different arbitrarily chosen nonlinear functions as examples for demonstrating the capabilities of the proposed architecture and its \emph{reconfigurability} in implementing nonlinear functions of the input signal.  
The first one is a quadratic function of the form $f(x) = x^2 / 4$. 
The target function of the input signal magnitude considering $x = A_\text{in} / A_\text{sat}$ is shown in Fig.~\ref{nonlinear_funcs}(a) (solid red line).
Assuming $N=5$, the differential phases of the five MZIs in the slope screen have been optimized to match the realized nonlinear output signal with the target quadratic function.
The blue solid line in Fig.~\ref{nonlinear_funcs}(a) shows the output signal obtained by our proposed architecture with the optimized slopes $ \textbf{s} = [0.42,0.61,0.56,0.49,0.87]$.
As observed in Fig.~\ref{nonlinear_funcs}(a), the agreement of the realized output signal with the target quadratic function is acceptable.
The same architecture can also implement, for instance, a third-degree polynomial function (Fig.~\ref{nonlinear_funcs}(b)), a sigmoid function (Fig.~\ref{nonlinear_funcs}(c)), and a transcendental function (Fig.~\ref{nonlinear_funcs}(d)), just by properly tuning the differential phases of the MZIs in the slope screen.
Also in these three examples, the realized output signal by the proposed architecture closely follows the target nonlinear functions.
Just like the quadratic function, the required differential phases of the MZIs in the slope screen for implementing the other three nonlinear functions have been obtained through an optimization process.
The resulting slopes are shown in the insets of Figs.~\ref{nonlinear_funcs}(b)-(d).
We point out that the optimization process to characterize the slope screen for implementing a nonlinear function needs to be run only once and usually takes little time (e.g., less than a second).
\section{Summary and Conclusion}
In this work, we theoretically proposed and numerically demonstrated a device that can exhibit reconfigurable nonlinear dependence between the input and output signals.  It consists of two networks, the first of which is a set of several identical inverse-designed three-port structures with an optimized mixture of linear and nonlinear materials, providing desired power limiting and power dividing characteristics.  The second network is a linear mesh of MZI elements that consists of two sections; the slope screen and the adder networks.  The reconfigurability of the MZI mesh enables one to change and tune at will the functional dependence of the entire device.  Several examples illustrating salient features of this device were given and discussed.  The fact that the proposed architecture is reconfigurable makes it useful for numerous applications such as optical neural networks and wave-based analog computing architectures with nonlinearity.    

\section*{Acknowledgements}
This work is supported in part by the US Air Force Office of Scientific Research (AFOSR) Multidisciplinary University Research Initiative (MURI) grant numbers FA9550-17-1-0002 and FA9550-21-1-0312, and in part by the US National Science
Foundation (NSF) MRSEC program under award No. DMR-1720530.
\bibliography{ms.bib}
\end{document}


\section*{Topology optimization procedure for the nonlinear element}
The distribution of $\text{As}_2 \text{S}_3$ and $\text{Si}_3 \text{N}_4$ within the design domain is interpolated by a density variable field $0\leq \rho (x,y) \leq 1$.
$\rho=1$ is projected to the permittivity of $\text{As}_2 \text{S}_3$, and the lower bound $\rho = 0$ to the permittivity of $\text{Si}_3 \text{N}_4$.
The intermediate values of the density, which have no physical meaning regarding the wave-matter interaction, must be penalized during the optimization.
In other words, the density must be appropriately binarized i.e., at every point, it only takes either zero or one.
Also, for controlling the feature sizes of the density distribution and removing small features for future fabrication consideration, a Helmholtz-type spatial filter is applied to the density which filters out domains that are smaller than $R_\text{min}$\cite{lazarov2011filters}:
\begin{equation}\tag{S1}
     \rho_\text{f} (x,y) - R_\text{min}^2 \nabla^2 \rho_\text{f} (x,y) = \rho (x,y) 
\end{equation}
where $\rho_\text{f} (x,y)$ is the filtered density as the solution to the above differential equation and $R_\text{min}$ is the filter radius. 
For penalizing the intermediate values and pushing towards a more binarized distribution, a projection function with a steepness factor $p$ is applied to the filtered density:
\begin{equation}\tag{S2}
    \rho_\text{p}= \frac{1}{1+e^{-p\left ( \rho_\text{f}- 0.5 \right )}}
\end{equation}
The steepness factor $p$ in the projection function determines the sharpness of the transition of $\rho_\text{f}$ from 0 to 1 around the middle value $\rho_\text{f}  = 0.5$.
The larger the $p$ parameter is, the more binarized the material distribution would be. During optimization iterations, the steepness factor becomes successively larger.
For material distribution, we used a linear interpolation scheme to map the projected density to the permittivity distribution inside the design region as follows:
\begin{equation}\tag{S3}
    \varepsilon_r (x,y,|E_z|^2) = \left ( 1 - \rho_\text{p} (x,y) \right ) \varepsilon_r^{\text{Si}_3 \text{N}_4} + \rho_\text{p} (x,y) \left ( \varepsilon_r^\text{L} + 3 \chi^{(3)} |E_z|^2 \right )
\end{equation}
Using the adjoint method for calculating the gradient, the optimization algorithm searches through the parameter space spanned by the density variable field to optimize the material distribution to achieve the desired response of the nonlinear power limiter.

For each target input power, $P_\text{in}$, we solve the frequency-domain Helmholtz's nonlinear wave equation as:
\begin{equation}\tag{S4}
    \nabla \times \left ( \mu_r^{-1} \nabla \times E_\text{z} \right ) -\omega_0^2 \mu_0 \varepsilon_0 \varepsilon_r \left ( x,y,|E_\text{z} |^2 \right )E_z = -j \omega_0 \mu_0 J_z
\end{equation}
Where $E_z $ is the $z$ component of the electric field throughout the structure, that is excited the current density $J_z$ at the input port boundary.
The current density generates an input signal as the fundamental waveguide mode that is injected with an input power $P_\text{in}$.

$\varepsilon_r \left ( x,y,|E_\text{z}|^2 \right )$ is the relative permittivity inside the simulation domain, and it is a function of magnitude of the electric field wherever the Kerr nonlinear material is present.
Finally, we set $\mu_0 = 4\pi \times 10^{-7} \ \text{H}/\text{m}$ ($\mu_r = 1$) all over the simulation domain.
To solve the above equation, we employ the Finite Element Method (FEM) together with Newton-Raphson iterative solver using commercial COMSOL Multiphysics\textsuperscript{\textregistered}\cite{COMSOL}.

For optimization, we set up a cost function to optimize the amount of power reaching Ports 2 and 3 depending on injected power through Port 1.
To do so, we define a splitting ratio, which is appropriately defined as a function of the input power, $R(P_\text{in})$ as following:
\begin{equation}\tag{S5}\label{split_ratio}
    \begin{split}
        & \displaystyle
        R(P_\text{in}) = 
        \begin{cases} 
            1, & P_\text{in} \leq P_\text{sat} \\
            P_\text{sat}/P_\text{in}, & P_\text{in} \geq P_\text{sat}  
        \end{cases} \\
        & P_2 = R(P_\text{in})P_\text{in} \\
        & P_3 = \left( 1-R(P_\text{in}) \right) P_\text{in}
    \end{split}
\end{equation}
Where $P_2$ and $P_3$ are the signal powers at Ports 2 and 3, respectively.
The defined splitting ratio implies that when the input power is below $P_\text{sat}$, almost all of it goes to Port 2 and almost none to Port 3.
When the input power is above $P_\text{sat}$ e.g., $P_\text{in}= A P_\text{sat}$, $A \geq 1$, the power at Port 2 will be $P_2 = RP_\text{in} = (1/A)(AP_\text{sat}) = P_\text{sat}$ satisfying the power saturation at Port 2, and the rest goes to Port 3.

The desired performance is to couple the input signal with power $P_\text{in}$ to Ports 2 and 3 with the splitting ratio defined in Eq.~(\ref{split_ratio}).
Therefore, we select $N$ target input powers,$P_\text{in}^k, \ \ k =1,2,...,N$ (shown as the red circles in Fig.~2(b) of the main text), and define the cost function according to:
\begin{equation}\tag{S6}
    \begin{split}
        L = \sum_{k=1}^{N} & \int_{\partial D} \left| E_\text{z}^\text{p2} - \sqrt{R(P_\text{in}^k) P_\text{in}^k} E^\text{m,p2}_z \right|^2 ds + \\
        & \int_{\partial D} \left| E_\text{z}^\text{p3} - \sqrt{\left(1-R(P_\text{in}^k)\right) P_\text{in}^k} E^\text{m,p3}_z \right|^2 ds
    \end{split}
\end{equation}
Where $E_\text{z}^\text{p2}$ and $E_\text{z}^\text{p3}$ are the simulated electric fields at Ports 2 and 3, respectively, at each iteration and corresponding to the target input power $P_\text{in}^k$.
$E_\text{z}^\text{m,p2}$ and $E_\text{z}^\text{m,p3}$ are the electric fields of the fundamental eigenmodes on Ports 2 and 3, respectively, that are normalized with respect to the input power $P_\text{in}^k$.
$\partial D$ denotes the waveguides' cross-sections covering the waveguides' cores and extends into the air cladding up to a point for covering the electric field profiles of the modes entirely.
Minimizing the above cost function gives the desired splitting ratio at target input powers.
If the target input powers are close enough to each other, the performance for other values of input powers also becomes close to the desired response.

In the range of input powers for which the element is designed, the local nonlinear response of the Kerr material might be weak.
Therefore one would need to enhance the collective power-dependent response of the wave-matter interaction within the design region.
There are two options for enhancing the collective power-dependent response.
The first is to design the material distribution to create an intense electromagnetic resonance.
A strong resonance would be sensitive to a slight change in the nonlinear refractive index due to the varying input power, thus giving us the desired strong collective power-dependent response.
However, this approach has the disadvantages of limiting the operational bandwidth and creating large-field hot spots, which would be problematic regarding material damage.
Furthermore, a strong resonance is sensitive to any unavoidable imperfection, including nanofabrication errors.

The second option for enhancing the collective intensity-dependent response is to manage for inverse design to achieve a distribution that mostly constitutes the Kerr material.
In that case, the EM wave would mostly travel inside the Kerr material to get to the output ports, hence a small modification of the refractive index of the Kerr material along with the long optical path of the wave traveling inside it would enhance the power-dependent response to saturate the power at Port 2.
We choose the latter approach for design.
One way to stir inverse design to fill most of the design region with Kerr material is simply initializing the density function such that at the start the design region is totally filled with Kerr material i.e., the initial density function is set to $\rho(x,y) = 1 $. 
As the optimization proceeds, the density function at some areas starts to deviate towards $\rho(x,y) = 0$ (corresponding to $\text{Si}_3 \text{N}_4$-filled holes) based on the computed gradient and after a few hundred iterations the optimized distribution is achieved.
\section*{Mathematical characterization of the MZI-based coherent adder}
As mentioned in the main text, the MZI-based adder utilized in the proposed nonlinear architecture  (Fig.~3(a) of the main text) combines the $N$ complex-valued signals leaving the slope screen into the output signal. To set the MZI phases of the adder, as suggested by Miller we can imagine running it backward by sending a signal from the output port \cite{miller2013self,MillerOptExp,MillerOptica}. The power injected from the output port needs to be equally divided and to appear at the adder's ``input'' ports with equal phase. This behavior can be mathematically described by a vector of entries of equal amplitudes and phases, such as \cite{MillerOptica}:
\begin{equation}\tag{S7}
        v_a = \frac{1}{\sqrt{N}}\left[1\right]_{\left(N-1\right) \times 1}
\end{equation}
Now, we can calculate the MZI phases of the adder, progressively from top to bottom, such that the signals appearing at the ``input'' ports are equal to $v_a$ \cite{miller2013self,MillerOptExp,MillerOptica}. Noticing that $v_a$ is a vector of real positive numbers, the common phase of each MZIs ($\phi$), which determines the phase of the output signals (see Eq.~(2) of the main text), can be set equal to $\pi/2$. The differential phase of the first MZI (the top one) can be derived by equating the first entry of $\textbf{T}_\text{MZI}$ and $v_a$, which results in $\theta_1=\arcsin\left(1/\sqrt{N}\right)$. To derive the differential phases of the subsequent MZIs, we need to take into account the propagation of the signal through the previous MZIs. For example, to characterize the second MZI, we can write $\sin\theta_2 \cos\theta_1=1/\sqrt{N}$ where the cosine, which comes from the second entry of $\textbf{T}_\text{MZI}$, describes the propagation of the signal through the first MZI. So, the differential phase of the second MZI is given by $\theta_2 = \arcsin\left(\frac{1}{\cos\theta_1 \sqrt{N}}\right)$. This process can be generalized to calculate the differential phase of any adder MZIs through the following expression: 
\begin{equation}\tag{S8}
       \theta_k = \arcsin\left( \frac{1}{\sqrt{N} \prod_{i=1}^{k-1} \cos\left(\theta_i\right)}\right)
\end{equation}
with $k=1,....,N-1$ and considering the product term equals $1$ for $k=1$. Note that, this characterization of the adder allows conveying all the power carried by the input signals to the output port only if the input signals are equal in amplitude and phase to $v_a$. In the context of the architecture of Fig.~3(a), as discussed in the main text, the signals have accumulated the same phase before entering the adder, but they have different amplitudes. Thus, the adder correctly adds algebraically the input signals to the output port (with a scale factor of $1/\sqrt{N}$) while not adding all the input power. The remaining power ends up in the bottom-right ports of the adder MZIs.

\bibliography{supplement.bib}